# SuperVoxHenry: Tucker-Enhanced and FFT-Accelerated Inductance Extraction for Voxelized Superconducting Structures

Mingyu Wang, *Graduate Student Member, IEEE*, Cheng Qian, Enrico Di Lorenzo, *Member, IEEE*, Luis J. Gomez, *Member, IEEE*, Vladimir Okhmatovski, *Senior Member, IEEE* and Abdulkadir C. Yucel, *Senior Member, IEEE*

*Abstract*— This paper introduces SuperVoxHenry, an inductance extraction simulator for analyzing voxelized superconducting structures. SuperVoxHenry extends the capabilities of the inductance extractor VoxHenry for analyzing the superconducting structures by incorporating the following enhancements. (i) SuperVoxHenry utilizes a two-fluid model to account for normal currents and supercurrents. (ii) SuperVoxHenry introduces the Tucker decompositions to reduce the memory requirement of circulant tensors as well as the setup time of the simulator. (iii) SuperVoxHenry incorporates an aggregation-based algebraic multigrid technique to obtain the sparse preconditioner. With these enhancements, SuperVoxHenry allows extracting the inductance of large-scale superconducting structures on a desktop computer. The accuracy, efficiency, and applicability of the proposed SuperVoxHenry have been demonstrated through the inductance extraction of various superconducting structures, including superconducting thin film inductors, a sharp bend, as well as a subsystem of an energy-efficient single flux quantum circuit.

## I. Introduction

The analysis and design of superconducting digital integrated circuits, including energy-efficient single flux quantum (eSFQ) circuits, Josephson junctions, inductors, and analog-to-digital converters, are crucial for the advancement in energy-efficient supercomputing and quantum technologies [1]-[3]. The design and verification of such circuits benefit from efficient and accurate inductance extraction simulators. So far, various inductance extraction simulators for 3D superconducting circuit models have been developed [4]-[16]. Among these, FastHenry [9], [12] still preserves its popularity. FastHenry allows analyzing the structures discretized by filaments. Another inductance extraction tool InductEx [6] extends FastHenry with efficient pre- and post-processors for complex multiport structures [7], [11]. As FastHenry employs filament-based discretization, it requires fine discretization around sharp corners, making it inefficient for the analysis of the structures with sharp bends and intricate geometrical features. Besides, the filament-based discretization makes the incorporation of ground planes in the analysis a highly tedious task for the users since 1D filaments are to be judiciously combined to model 2D/3D currents on the ground planes, as described in [17]. The same problems persist in the multilevel adaptive cross approximation accelerated inductance extraction simulator in [14] as it also leverages filaments. To overcome these problems associated with the filament-based discretization, TetraHenry was developed to analyze superconducting structures discretized by tetrahedral elements [10]. However, TetraHenry and all the abovementioned inductance extraction simulators are based on mesh formulation [12]. They use loop basis functions and require the search for the loops, necessitating excessive computational resources when analyzing structures on complicated ground planes with holes. Furthermore, none of these simulators were specifically developed for inductance extraction of the voxel-discretized structures and thereby cannot provide the inductances of voxelized structures with maximum possible efficiency.

Recently, an inductance extraction simulator called VoxHenry has been proposed [18]. The VoxHenry leverages voxel-based discretization. Thereby, it is free of the deficiencies resulting from filament-based discretization. It uses the piecewise constant and linear basis functions to model currents inside voxels. By doing so, it can accurately model the currents turning the corners. Therefore, it does not require any special treatment for the corners and ground planes. Furthermore, VoxHenry is based on nodal formulation [12] and does not require any search for the loops as in the mesh-based approach. Consequently, it can efficiently perform the analysis of structures on complicated ground planes with holes.

In this study, we extend the VoxHenry simulator for the analysis of voxelized superconducting structures. The proposed open-source simulator, SuperVoxHenry, solves the volume integral equation (VIE) along with current conservation equation after accounting for the two-fluid model of normal currents and supercurrents. SuperVoxHenry exploits fast Fourier transforms (FFTs) to expedite the matrix-vector

Manuscript received April 23, 2021. This work was supported by Ministry of Education, Singapore, under grant AcRF TIER 1-2018-T1-002-077 (RG 176/18), and the Nanyang Technological University under a Start-Up Grant. (*Corresponding author: Abdulkadir C. Yucel.*)

M. Wang, C. Qian and A. C. Yucel are with the School of Electrical and Electronic Engineering, Nanyang Technological University, Singapore 639798. (e-mails: mingyu003@e.ntu.edu.sg, cqian@ntu.edu.sg, acyucel@ntu.edu.sg).

Enrico Di Lorenzo is with FastFieldSolvers S.R.L.Vimercate 20871, Italy. (e-mail: enrico.di_lorenzo@fastfieldsolvers.com)

L. J. Gomez is with the Department of Electrical and Computer Engineering, Purdue University, West Lafayette, IN 47907 USA. (e-mail: ljgomez@purdue.edu).

Vladimir Okhmatovski is with Dept. of ECE, University of Manitoba, Winnipeg, MB R3T 2N2, CA. (e-mail: Vladimir.Okhmatovski@umanitoba.ca)

multiplications during the iterative solution of the discretized governing equations. To ensure the iterative solution's fast convergence, the SuperVoxHenry uses a sparse preconditioner, obtained by an aggregation-based algebraic multigrid (shortly AGMG) method [19], [20]. AGMG method efficiently inverts the Schur complement of the sparse preconditioner and drastically reduces the sparse preconditioner's memory requirement. In addition, SuperVoxHenry employs the Tucker decompositions [21], [22] to reduce the memory requirement of the circulant tensors as well as the CPU time required for the setup stage of the simulator. By considering these features, the contributions of this paper are claimed as three-fold. First, it extends the VoxHenry simulator to the inductance extraction of the voxelized superconducting structures. It provides the details of the modifications performed on VoxHenry's formulation. Second, it proposes using the AGMG method to invert the Schur complement of the sparse preconditioner iteratively. By doing so, direct inversion of the Schur complement via LDLT decomposition in [18], requiring more than half of the memory requirement of the simulator, is avoided. Third, it proposes Tucker decompositions to compress the system tensors generated for the magneto-quasi-static analysis for the first time. In the past, Tucker decompositions were used to compress the system tensors generated for the electrostatic [23] and time-harmonic [24]-[28] analyses, yet have never been used before for compressing the system tensors in the magneto-quasi-static analysis.

Tucker decompositions allow a dramatic reduction in the simulator's setup time and the circulant tensors' memory requirement via the strategy explained as follows. During the simulator's installation stage, the Tucker-compressed Toeplitz tensors are generated for a large computational domain and stored on the hard disk. During the setup stage of the simulator's each execution, the compressed tensors are read from the hard disk and restored to their original formats. The restored tensors are resized according to the size of the analyzed structure's computational domain and multiplied by a scaling factor. The restored Toeplitz tensors are then used to obtain the circulant tensors via embedding. By doing so, the CPU time required to generate Toeplitz and circulant tensors is reduced from thousands of seconds to tens of seconds. Note that this reduction leads to a substantial reduction in the CPU time required by the simulator's setup stage. Moreover, the circulant tensors are compressed via Tucker decomposition during the setup stage and restored one-by-one during the iterative solution stage. By doing so, the memory requirement of the circulant tensors is reduced from tens of gigabytes to a few megabytes. Such memory saving gives rise to a significant reduction in the memory footprint of the SuperVoxHenry simulator. Note that one-by-one restoration of the circulant tensors during iterative solution requires negligible computational overhead.

The proposed SuperVoxHenry simulator is applied to inductance extraction of various structures, including superconducting microstrip, inverted microstrip, and striplines inductors [29], 90-degree bend, and 10-bit eSFQ register [1]. Some important observations on its application are in order: 1) The proposed SuperVoxHenry outperforms the FastHenry for the analyses of various voxelized superconducting structures. For example, SuperVoxHenry requires 13.13x less memory and 876.20x less CPU time than FastHenry to achieve the same level of accuracy for the inductance extraction of 90-degree sharp bend. 2) The Tucker enhancement significantly reduces the CPU time for obtaining Toeplitz tensors and lessens the circulant tensors' memory requirement. For example, for analyzing a voxelized superconducting structure discretized by 500 voxels along each principal axis, the Tucker decomposition reduces the CPU time for obtaining Toeplitz tensors by a factor of 43, while it lessens the memory requirement of the circulant tensors by more than 20,000x. 3) The scaling of the memory requirement of the Tucker-compressed tensors with respect to computational domain size is found to be sub-linear, while that of the memory requirement of the original tensors are linear. For example, Tucker-compressed circulant tensors' memory requirement scales as $O(K_t^{0.25} \log K_t)$, while the memory scaling of original circulant tensors is $O(K_t)$, where $K_t$ is the total number of voxels in the computational domain. 4) The AGMG method requires 12x and 5x less peak memory compared to the LDLT method for inductance extraction of the sharp bend and 10-bit eSFQ register, respectively.

## II. FORMULATION

In this section, the formulation of the SuperVoxHenry for inductance extraction of the superconducting structures is provided. To this end, a similar notation in [18] is adopted here. Next, Tucker decomposition methodology for reducing the memory requirement of the circulant tensors as well as the CPU time of the setup stage is expounded. Finally, the AGMG method's application for reducing the memory requirement of the sparse preconditioner is explained.

*A. The Discretization and Accelerated Solution of Governing Equations in SuperVoxHenry*

Let $V'$ represent a volume comprising one or more superconductors with permeability $\mu = \mu_0$; $\mu_0$ is the permeability of the free space. The superconductor(s) is/are connected to a sinusoidal voltage source operated at an angular frequency $\omega$. The structure is enclosed by a computational domain, which is discretized by $K_t$ voxels with an edge length $\Delta x$. Along each dimension of the coordinate system, there are $K_x$, $K_y$, and $K_z$ number of voxels and $K_t = K_x \times K_y \times K_z$. There exist $K$ voxels occupied by the superconductors, referred to as non-empty voxels; the others are named as empty voxels. Each non-empty voxel has six nodes defined on the centers of its faces. When the voxel has non-empty neighbors, some nodes are shared. The total number of unique nodes is $M$. For this voxelized setting, SuperVoxHenry solves the VIE and current conservation law [12, 30], which are

$$\frac{\mathbf{J}(\mathbf{r})}{\sigma(\mathbf{r})} + j\omega\mu \int_{V'} G(\mathbf{r},\mathbf{r}') \mathbf{J}(\mathbf{r}') dv' = -\nabla \Phi(\mathbf{r}), \quad (1)$$

$$\nabla \cdot \mathbf{J}(\mathbf{r}) = 0, \quad (2)$$

where, $G(\mathbf{r},\mathbf{r}') = 1/(4\pi|\mathbf{r}-\mathbf{r}'|)$ is the free-space Green's function, $\mathbf{r}$ and $\mathbf{r}'$ represent observation and source locations in $V'$, $\mathbf{J}(\mathbf{r})$ and $\Phi(\mathbf{r})$ denote the unknown vector current density and the scalar potential, respectively. Here $\sigma(\mathbf{r})$ is the

two-fluid conductivity, defined as [9, 10, 14, 31]

$$\sigma(\mathbf{r}) = \sigma_0(\mathbf{r}) - j/(\omega\mu\lambda(\mathbf{r})^2). \quad (3)$$

Here $\sigma_0(\mathbf{r})$ is the temperature-dependent conductivity of the normal channel and $\lambda(\mathbf{r})$ is the London penetration depth. (3) allows modeling normal and superconducting channels. The currents in these channels are discretized by a set of divergence-free basis functions defined in each non-empty voxel as [18]

$$\mathbf{J}(\mathbf{r}) \approx \sum_{k=1}^{K} I_k^x \mathbf{f}_k^x(\mathbf{r}) + I_k^y \mathbf{f}_k^y(\mathbf{r}) + I_k^z \mathbf{f}_k^z(\mathbf{r}) \\ + I_k^{2D} \mathbf{f}_k^{2D}(\mathbf{r}) + I_k^{3D} \mathbf{f}_k^{3D}(\mathbf{r}), \quad (4)$$

where $\mathbf{f}_k^x(\mathbf{r}) = \hat{\mathbf{x}}$, $\mathbf{f}_k^y(\mathbf{r}) = \hat{\mathbf{y}}$, and $\mathbf{f}_k^z(\mathbf{r}) = \hat{\mathbf{z}}$ are the piecewise constant basis functions; $\mathbf{f}_k^{2D}(\mathbf{r})$ and $\mathbf{f}_k^{3D}(\mathbf{r})$ are the piecewise linear basis functions, defined as

$$\mathbf{f}_k^{2D}(\mathbf{r}) = \left((x-x_k)\hat{\mathbf{x}} - (y-y_k)\hat{\mathbf{y}}\right)/\Delta x, \quad (5)$$

$$\mathbf{f}_k^{3D}(\mathbf{r}) = \left((x-x_k)\hat{\mathbf{x}} + (y-y_k)\hat{\mathbf{y}} - 2(z-z_k)\hat{\mathbf{z}}\right)/\Delta x, \quad (6)$$

Here, $(x_k, y_k, z_k)$ are the coordinates of the $k^{th}$ non-empty voxel's center. In (4), $I_k^x$, $I_k^y$, $I_k^z$, $I_k^{2D}$, and $I_k^{3D}$ are the unknown coefficients associated with each basis function, $k=1,\ldots,K$. Substituting the discretized current density into (1) and (2), applying Galerkin testing, and enforcing the current continuity on $M$ nodes yields a $N$ by $N$ linear system of equations, $N = 5K + M$, as

$$\begin{bmatrix} \mathbf{V} \\ 0 \end{bmatrix} = \begin{bmatrix} \bar{\mathbf{Z}} & -\bar{\mathbf{A}}^T \\ \bar{\mathbf{A}} & 0 \end{bmatrix} \begin{bmatrix} \mathbf{I} \\ \mathbf{\Phi} \end{bmatrix}, \quad (7)$$

where $\mathbf{I} = [\mathbf{I}^x; \mathbf{I}^y; \mathbf{I}^z; \mathbf{I}^{2D}; \mathbf{I}^{3D}]$ is the unknown current coefficient vector. The sparse matrix $\bar{\mathbf{A}}$ is the incidence matrix while $\mathbf{\Phi}$ and $\mathbf{V}$ are the potential coefficient vector and excitation vector, respectively. The entries of $\bar{\mathbf{A}}$, $\mathbf{\Phi}$, $\mathbf{V}$, and $\mathbf{I}$ are provided in [18]. The matrix $\bar{\mathbf{Z}}$ can be written explicitly as [18]

$$\bar{\mathbf{Z}} = \begin{bmatrix} \bar{\mathbf{Z}}^{x,x} & 0 & 0 & \bar{\mathbf{Z}}^{x,2D} & \bar{\mathbf{Z}}^{x,3D} \\ 0 & \bar{\mathbf{Z}}^{y,y} & 0 & \bar{\mathbf{Z}}^{y,2D} & \bar{\mathbf{Z}}^{y,3D} \\ 0 & 0 & \bar{\mathbf{Z}}^{z,z} & 0 & \bar{\mathbf{Z}}^{z,3D} \\ \bar{\mathbf{Z}}^{2D,x} & \bar{\mathbf{Z}}^{2D,y} & 0 & \bar{\mathbf{Z}}^{2D,2D} & \bar{\mathbf{Z}}^{2D,3D} \\ \bar{\mathbf{Z}}^{3D,x} & \bar{\mathbf{Z}}^{3D,y} & \bar{\mathbf{Z}}^{3D,z} & \bar{\mathbf{Z}}^{3D,2D} & \bar{\mathbf{Z}}^{3D,3D} \end{bmatrix}. \quad (8)$$

The entries of the blocks in $\bar{\mathbf{Z}}$, represented as $\bar{\mathbf{Z}}^{\beta,\alpha}$, $\beta,\alpha \in \{x,y,z,2D,3D\}$, are provided in [18]. To incorporate the two-fluid model, the entries of the diagonal blocks provided in [18] are modified as shown in Appendix A.

During the iterative solution of (7), the most time-consuming operation is the multiplications of the full blocks $\bar{\mathbf{Z}}^{\beta,\alpha}$, $\beta,\alpha \in \{x,y,z,2D,3D\}$, with pertinent current coefficient vectors $\mathbf{I}^\alpha$. The computational and memory costs of this operation are $O(K^2)$. The computational and memory costs can be reduced to $O(K_t \log K_t)$ and $O(K_t)$ via the FFT acceleration procedure. In this procedure, the block Toeplitz tensors $\mathcal{T}^{\beta,\alpha}$ with dimensions $K_x \times K_y \times K_z$, corresponding to the blocks $\bar{\mathbf{Z}}^{\beta,\alpha}$, $\beta,\alpha \in \{x,y,z,2D,3D\}$ are obtained first. Next, circulant tensors $\mathcal{Z}^{\beta,\alpha}$ with dimensions $2K_x \times 2K_y \times 2K_z$ are obtained from Toeplitz tensors $\mathcal{T}^{\beta,\alpha}$ by applying the embedding procedure [18] and computing the FFT of the resulting tensor. The matrix-vector multiplications $\mathbf{CC}^\beta = \bar{\mathbf{Z}}^{\beta,\alpha} \mathbf{I}^\alpha$ can be executed by FFTs

$$\mathcal{CC}^\beta = z^\beta \mathcal{I}^\beta + IFFT\left\{\sum_\alpha \mathcal{Z}^{\beta,\alpha} * \breve{\mathcal{I}}^\alpha\right\}. \quad (9)$$

Here, $\breve{\mathcal{I}}^\alpha = FFT\{\mathcal{I}^\alpha\}$. $*$, $FFT$, and $IFFT$ denote tensor-tensor multiplication, FFT, and inverse FFT operators, respectively. $\mathcal{I}^\alpha$ is filled by the entries of $\mathbf{I}^\alpha$ and zeros. $z^\beta$ is $c(\mathbf{r})/\Delta x$ for $\beta = \{x,y,z\}$, $c(\mathbf{r})/6\Delta x$ for $\beta = 2D$, and $c(\mathbf{r})/2\Delta x$ for $\beta = 3D$, where $c(\mathbf{r}) = a(\mathbf{r}) + j\omega\mu b(\mathbf{r})$; $a(\mathbf{r})$ and $b(\mathbf{r})$ are defined in Appendix A. After the multiplication, the result $\mathbf{CC}^\beta$ is obtained from the entries of $\mathcal{CC}^\beta$. Note that $\sigma(\mathbf{r})$ is embedded in $a(\mathbf{r})$ and $b(\mathbf{r})$. Therefore, the incorporation of the two-fluid model does not affect FFT and IFFT operations.

### B. Tucker Decomposition

The Toeplitz and circulant tensors, $\mathcal{T}^{\beta,\alpha}$ and $\mathcal{Z}^{\beta,\alpha}$, $\beta,\alpha \in \{x,y,z,2D,3D\}$, are low-rank tensors and can be compressed by Tucker decomposition. Let $\mathcal{O}$ represent a tensor ($\mathcal{T}^{\beta,\alpha}$ or $\mathcal{Z}^{\beta,\alpha}$) with dimensions $D_1 \times D_2 \times D_3$. The Tucker representation of this tensor is [21, 25, 32, 33]

$$\mathcal{O} = \mathcal{C} \times_1 \bar{\mathbf{F}}^1 \times_2 \bar{\mathbf{F}}^2 \times_3 \bar{\mathbf{F}}^3, \quad (10)$$

where $\mathcal{C}$ is the core tensor with dimensions $d_1 \times d_2 \times d_3$. $\bar{\mathbf{F}}^i$, $i=1,2,3$, represent the factor matrices with dimensions $D_i \times d_i$, and $d_i$ is the multilinear rank pertinent to $i^{th}$ dimension. As $\mathcal{T}^{\beta,\alpha}$ and $\mathcal{Z}^{\beta,\alpha}$ are low-rank, $(d_1 d_2 d_3) + \sum_{i=1}^{3} D_i d_i \ll D_1 D_2 D_3$ and these tensors are Tucker compressible. The symbol $\times_i$ in (10) denotes $i-$mode multiplication, performed for example for the $1^{st}$ dimension as

$$\mathcal{V}_{v_1,v_2,v_3} = \mathcal{C} \times_1 \bar{\mathbf{F}}^1 = \sum_{i'=1}^{d_1} \mathcal{C}_{i',v_2,v_3} \bar{\mathbf{F}}^1_{v_1,i'}, \quad (11)$$

where $\mathcal{V}$ is the resulting tensor with indices $v_1 = 1,\ldots,D_1$, $v_2 = 1,\ldots,d_2$, and $v_3 = 1,\ldots,d_3$. To compute the core tensor and factor matrices of the Tucker representation, algorithms based on singular value decomposition (SVD) [21] or cross approximation [34] can be leveraged. Here SVD-based method is briefly explained in Algorithm 1 for the sake of completeness. This algorithm is executed for a given tensor $\mathcal{O}$ and tolerance $tol$ to truncate the SVDs of the unfolding matrices, obtained by reshaping operations [33]. In this study, $tol$ is set to $10^{-8}$ to avoid sacrificing from accuracy while compressing, unless stated otherwise.

*1) Compressing the Toeplitz tensors:* The compression of the Toeplitz tensors allows accelerating the setup stage of the simulator. To do that, a strategy outlined in Algorithm 2 is executed. In this strategy, during the installation stage of the simulator, all $\mathcal{T}^{\beta,\alpha}$, $\beta,\alpha \in \{x,y,z,2D,3D\}$, are computed for a very large computational domain, compressed via the Tucker decomposition, and then stored in the hard disk. (Note: The size of the computational domain is determined as the maximum possible grid of the largest possible structure that can be simulated with the available RAM of the machine.) Note that the memory requirement of $\mathcal{T}^{\beta,\alpha}$ is often at the orders of gigabytes, while that of Tucker-compressed $\mathcal{T}^{\beta,\alpha}$ is around megabytes. During the setup stage of the simulator's each execution, the compressed representations are read from the hard disk and used to restore the original tensors via (10). The original tensors are then trimmed for the given computational domain with $K_x$, $K_y$, and $K_z$. Finally, the entries of the resulting tensors are multiplied by the scaling factor $(\Delta x)^5$; The derivation of the scaling factor is provided in Appendix B. The computational saving achieved by implementing the proposed strategy is demonstrated in Table II of Section III.B.3.

*2) Compressing the circulant tensors:* The compression of the circulant tensors allows reducing the memory footprint of the simulator. To do that, the circulant tensors $\mathcal{Z}^{\beta,\alpha}$, $\beta,\alpha \in \{x,y,z,2D,3D\}$, are compressed by Tucker decomposition using Algorithm 1. The Tucker-compressed tensors are restored to their original format *one-by-one* and used in (9) to perform the convolutions during the iterative solution stage of the simulator. Note that the restoration performed via (10) requires negligible computational overhead compared to the FFT-accelerated convolution in (9). The computational saving achieved by and the computational overhead introduced by compressing the circulant tensors are demonstrated in Section III.B.1 and Section III.B.2.

---

**Algorithm 1**: Tucker-SVD

1: **Inputs:** 3-D array $\mathcal{O} = \{\mathcal{T}^{\beta,\alpha}, \mathcal{Z}^{\beta,\alpha}\}$ and $tol$.
2: **Outputs:** $\mathcal{C}$ and $\bar{\mathbf{F}}^i$, $i=1,\ldots,3$.
3: **Initialize:** $\mathcal{C} = \mathcal{O}$.
4: **for** $i = 1:3$ **do**
5:    obtain mode $-i$ unfolding matrix of $\mathcal{O}$, $\bar{\mathbf{O}}^i$.
6:    obtain SVD of $\bar{\mathbf{O}}^i$ as $\bar{\mathbf{O}}^i = \bar{\mathbf{U}}^i \bar{\mathbf{\Sigma}}^i \bar{\mathbf{V}}^{i*}$.
7:    assign the index of maximum (normalized) value in $\bar{\mathbf{\Sigma}}^i$ smaller than $tol/\sqrt{3}$ as $d_i$, truncate $\bar{\mathbf{U}}^i$ with $d_i$, and assign $\bar{\mathbf{F}}^i = \bar{\mathbf{U}}^i$.
8:    $\mathcal{C} = \mathcal{C} \times_i \bar{\mathbf{F}}^{i*}$.
9: **end for**
10: **return**

---

**Algorithm 2**: Compression of $\mathcal{T}^{\beta,\alpha}$ for accelerating setup

1: **Pre-processing:** (during installation)
2: **for** loop over all Toeplitz tensors **do**
3:    obtain $\tilde{\mathcal{T}}^{\beta,\alpha}$ for a large domain (e.g., $K_x = K_y = K_z = 500$) by setting $\Delta x = 1$
4:    use Algorithm 1 to obtain Tucker core $\tilde{\mathcal{C}}^{\beta,\alpha}$ and factor matrices $\tilde{\bar{\mathbf{F}}}^{i,\beta,\alpha}$ of each $\tilde{\mathcal{T}}^{\beta,\alpha}$ and store in hard disk
5: **end for**
6: **Execution:** (during setup stage)
7: **Inputs:** $K_x$, $K_y$, and $K_z$
8: **Outputs:** $\mathcal{T}^{\beta,\alpha}$.
9: Read $\tilde{\mathcal{C}}^{\beta,\alpha}$ and $\tilde{\bar{\mathbf{F}}}^{i,\beta,\alpha}$ from hard disk
10: **for** loop over all Toeplitz tensors **do**
11:    obtain $\tilde{\mathcal{T}}^{\beta,\alpha}$ from $\tilde{\mathcal{C}}^{\beta,\alpha}$ and $\tilde{\bar{\mathbf{F}}}^{i,\beta,\alpha}$ via (10)
12:    set $\mathcal{T}^{\beta,\alpha} = \tilde{\mathcal{T}}^{\beta,\alpha}_{1:K_x, 1:K_y, 1:K_z}$
13:    multiply $\mathcal{T}^{\beta,\alpha}$ with scaling factor $(\Delta x)^5$
14: **end for**
15: **return**

---

*C. Sparse Preconditioner and Inversion of Schur Complement via AGMG Method*

The fast convergence of the iterative solution of (7) can be ensured by a sparse preconditioner defined as [18]

$$\bar{\mathbf{P}} = \bar{\mathbf{Q}}^{-1} = \begin{bmatrix} \bar{\mathbf{Y}} & -\bar{\mathbf{A}}^T \\ \bar{\mathbf{A}} & 0 \end{bmatrix}^{-1}, \quad (12)$$

where $\bar{\mathbf{Y}}$ is a diagonal matrix filled with the magnitudes of diagonal entries of $\bar{\mathbf{Z}}$. The inversion of $\bar{\mathbf{Q}}$ can be performed by applying the Schur complement. Assume that the multiplication of $\bar{\mathbf{P}}$ with a column vector $[\mathbf{c};\mathbf{d}]$ yields a column vector $[\mathbf{a};\mathbf{b}]$, i.e.,

$$\begin{bmatrix} \mathbf{a} \\ \mathbf{b} \end{bmatrix} = \begin{bmatrix} \bar{\mathbf{Y}} & -\bar{\mathbf{A}}^T \\ \bar{\mathbf{A}} & 0 \end{bmatrix}^{-1} \begin{bmatrix} \mathbf{c} \\ \mathbf{d} \end{bmatrix}. \quad (13)$$

To obtain $[\mathbf{a};\mathbf{b}]$, first, $\mathbf{b}$ is obtained as

$$\mathbf{b} = \bar{\mathbf{S}}^{-1}\mathbf{e}, \quad (14)$$

where $\bar{\mathbf{S}} = \bar{\mathbf{A}}\bar{\mathbf{Y}}^{-1}\bar{\mathbf{A}}^T$ is the Schur complement and $\mathbf{e} = \mathbf{d} - \bar{\mathbf{A}}\bar{\mathbf{Y}}^{-1}\mathbf{c}$. Second, $\mathbf{a}$ is obtained using $\mathbf{b}$ as

$$\mathbf{a} = \bar{\mathbf{Y}}^{-1}(\mathbf{c} + \bar{\mathbf{A}}^T\mathbf{b}). \quad (15)$$

Inversion of $\bar{\mathbf{Y}}$ is trivial as it is a diagonal matrix. On the other hand, inversion of Schur complement $\bar{\mathbf{S}}$ can be performed via LDLT decomposition in [15] as it is sparse, positive-definite, and symmetric real matrix. However, the memory requirement of the LDLT decomposition is nearly half of the simulator's memory footprint. Here we propose to use the AGMG method [19] to iteratively invert the Schur complement. The AGMG method was also used to invert the Schur complement in [35] for a linear system and Schur complement different from those in our study. The open-source AGMG method, provided in [20], is a black-box inversion technique, which takes $\bar{\mathbf{S}}$ and $\mathbf{e}$ in (14) as inputs and provides $\mathbf{b}$ as output for a given tolerance, set to $10^{-8}$ in this study. The method uses conjugate gradient method for iterative solution and accelerates the convergence with the algebraic multigrid preconditioner. The memory requirement for the AGMG-based iterative inversion is significantly less compared to the memory requirement of Schur complement's inversion via LDLT decomposition, as shown in Sections III.C and III.D. Furthermore, as the AGMG performs only a handful of iterations during the inversion, the computational overhead compared to LDLT decomposition is acceptable, as discussed in Sections III.C and III.D.

## III. NUMERICAL RESULTS

This section presents several numerical examples demonstrating the accuracy, efficiency, and applicability of the SuperVoxHenry simulator. First, the SuperVoxHenry's accuracy is validated by comparing the inductances obtained by SuperVoxHenry and those obtained by FastHenry and measurements. Next, the memory saving and computational overhead introduced by compressing circulant tensors as well as the memory and computational savings obtained by compressing Toeplitz tensors are extensively quantified. Finally, the proposed simulator is applied to the inductance extraction of a 90-degree sharp bend and a 10-bit eSFQ shift register. When applicable, the relative difference between inductances obtained by SuperVoxHenry and FastHenry is quantified via $err = |\tilde{I} - I|/|\tilde{I}|$, where $\tilde{I}$ and $I$ denote the inductances obtained by SuperVoxHenry and FastHenry, respectively.

It should be noted that both simulators are executed for the same voxelized structures with the same discretizations for fair comparisons in the tests below. However, FastHenry employs one-fifth of the basis functions employed by SuperVoxHenry. Thereby, the number of unknowns solved by FastHenry is far less than those solved by SuperVoxHenry. The proposed open-source SuperVoxHenry simulator is implemented in Matlab, while the FastHenry simulator executes a C code. Both simulators solve their linear system of equations iteratively by a generalized minimal residual (GMRES) method with a restart of every 50 iterations until a relative residual error (RRE) of $10^{-8}$ is achieved, unless stated otherwise. All simulations are carried out on an Intel Xeon Gold 6412 CPU with 384 GB RAM.

### A. Validation Examples

The SuperVoxHenry is used to extract inductances of superconducting inductors proposed in [29]. These structures include microstrip, inverted microstrip, and stripline thin-film inductors, realized by metal layers M6-M4, M6-M7, and M5-M6-M7 with the names provided in [29] and sectional views shown in the inlets of Figs. 1(a), (b), and (c), respectively. In these structures [Figs.1(a)-(c)], the dimensions of ground planes (GND) are $5\times0.2\times10\,\mu m$ (width x height x length) while those of the strips (Signal) are $w\times0.2\times10\,\mu m$, where $w$ is the width of the strip. The spacing between the ground plane and the strip is $0.2\,\mu m$ for inverted microstrip and stripline, while it is $0.6\,\mu m$ for microstrip. The structures are discretized by voxels with $\Delta x = 0.05\,\mu m$. For increasing $w$, the structures' per-unit-length inductances are computed via the proposed SuperVoxHenry by setting, $\lambda = 9\times10^{-8}$ m, and $\sigma_0 = 0$. The computed results are compared with the measurement results as well as the ones obtained by FastHenry simulator with $2^{nd}$-order multipole expansion [Fig. 1(a)-(c)]. Needless to say, all results are in very good agreement. The maximum and minimum relative differences between the SuperVoxHenry and FastHenry results are $err = 0.022$ and $err = 0.009$, respectively. Furthermore, the current distributions on microstrip and stripline are plotted when the strip width is $2\,\mu m$ [Fig. 1(d)]. For this example, the specifications of the simulations performed by SuperVoxHenry and FastHenry for $w = 2\,\mu m$ are provided in Table I. For these structures, the Tucker enhancement reduces SuperVoxHenry's memory requirement by a factor changing from 1.26 to 1.86. At the same time, it requires a computational overhead changing from 2.13% to 4.88% (see the definition of computational overhead in Section III.B). In addition, the SuperVoxHenry requires 4.38x (min) and 5.79x (max) less memory as well as 19.87x (min) and 27.05x (max) less CPU time compared to the FastHenry.

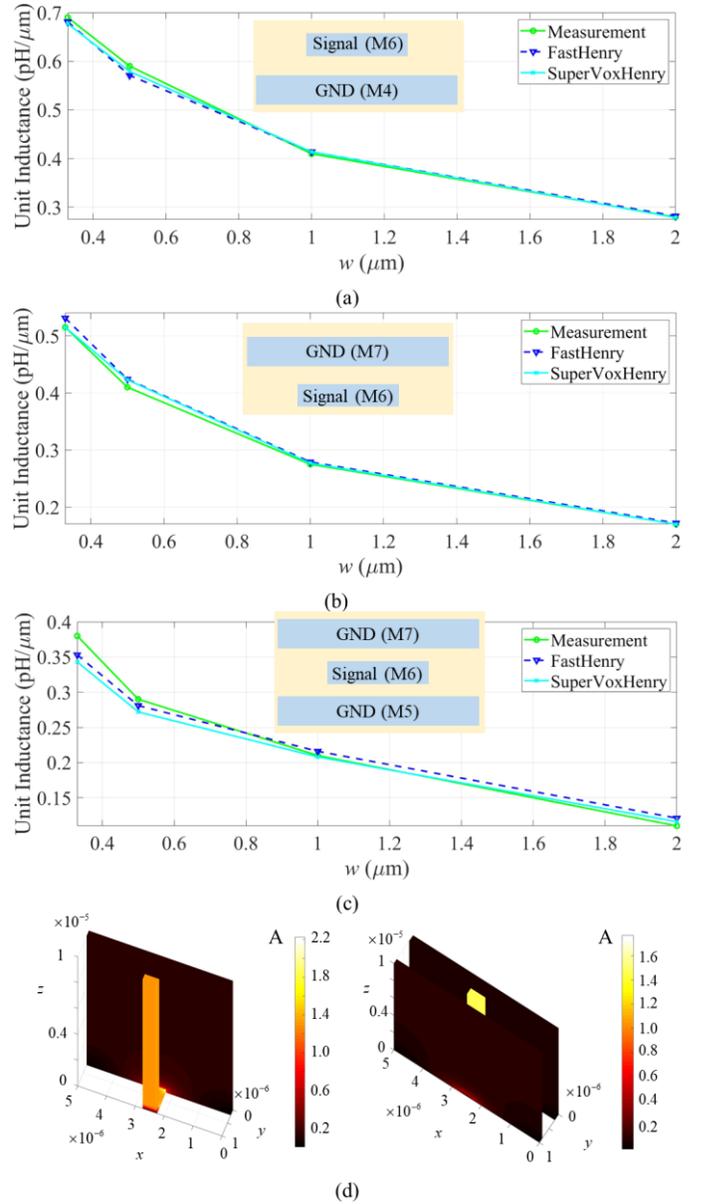

Fig. 1. Per-unit-length inductance of superconducting inductors realized by metal layers (a) M6-M4 (microstrip) (b) M6-M7 (inverted microstrip) (c) M5-M6-M7 (stripline), and the current distributions on (d) microstrip (left) and stripline (right).

TABLE I
COMPUTATIONAL REQUIREMENTS OF THE SIMULATIONS PERFORMED BY SUPERVOXHENRY AND FASTHENRY FOR SUPERCONDUCTING INDUCTORS

| Structure | Simulator | Memory (MB) | CPU time (s) | Number of unknowns |
|---|---|---|---|---|
| Stripline (M5-M6-M7) | SuperVoxHenry w/o Tucker | 2,400 | 183.69 | 1,399,752 |
| | SuperVoxHenry w/ Tucker | 1,900 | 187.97 | 1,399,752 |
| | FastHenry | 11,000 | 3,735.41 | 170,436 |
| Microstrip (M6-M4) | SuperVoxHenry w/o Tucker | 2,400 | 118.08 | 941,888 |
| | SuperVoxHenry w/ Tucker | 1,300 | 120.59 | 941,888 |
| | FastHenry | 5,700 | 2,906.59 | 114,638 |
| Microstrip (M6-M7) | SuperVoxHenry w/o Tucker | 1,500 | 110.5 | 931,296 |
| | SuperVoxHenry w/ Tucker | 1,000 | 115.89 | 931,296 |
| | FastHenry | 5,700 | 2,989.79 | 113,358 |

### B. Tucker Enhancement

Next, the performance of the Tucker enhancement is extensively quantified. For this purpose, the compression ratio (CR), memory saving achieved by compressing, is defined as the ratio of the original tensors' memory requirement to the compressed tensors' memory requirement. In addition, computational overhead (CO), required for restoring the tensor, is defined as the ratio of the restoration time to the time required for one convolution performed with the same tensor.

*1) Compression of circulant tensors:* First, a superconducting square plate with a thickness of $3\,\mu m$ [Fig. 2(a)] is considered to demonstrate the performance of Tucker enhancement on 2D-like structures. The square plate's edge length is set to $\{300, 400, 700, 900, 1000, 1400\}\,\mu m$. For each different edge length, the plate is discretized by voxels of size $\Delta x = 1\,\mu m$; the generated circulant tensors are compressed by Tucker decomposition. The resulting CR and CO are plotted in Figs. 2(a) and (b), respectively. Apparently, CR increases with increasing $tol$ and $K_t$; it reaches above 700 (for $tol = 10^{-4}$) for the largest structure. Furthermore, CO decreases with increasing $K_t$ and became less than 0.3 for the largest structure, which shows that the computational overhead is negligible. Increase in CR and decrease in CO with increasing $K_t$ result from the fact that the maximum multilinear rank remains nearly constant with increasing $K_t$. This fact is clearly demonstrated in Fig. 2(c) where the maximum multilinear ranks of block $\mathcal{Z}^{x,x}$ for various $tol$ values are plotted with respect to $K_t$. Moreover, the original tensors' memory requirement scales as $O(K_t)$, while the compressed tensors' memory requirement scales as $O(K_t^{0.49}\log K_t)$ [Fig. 2(d)].

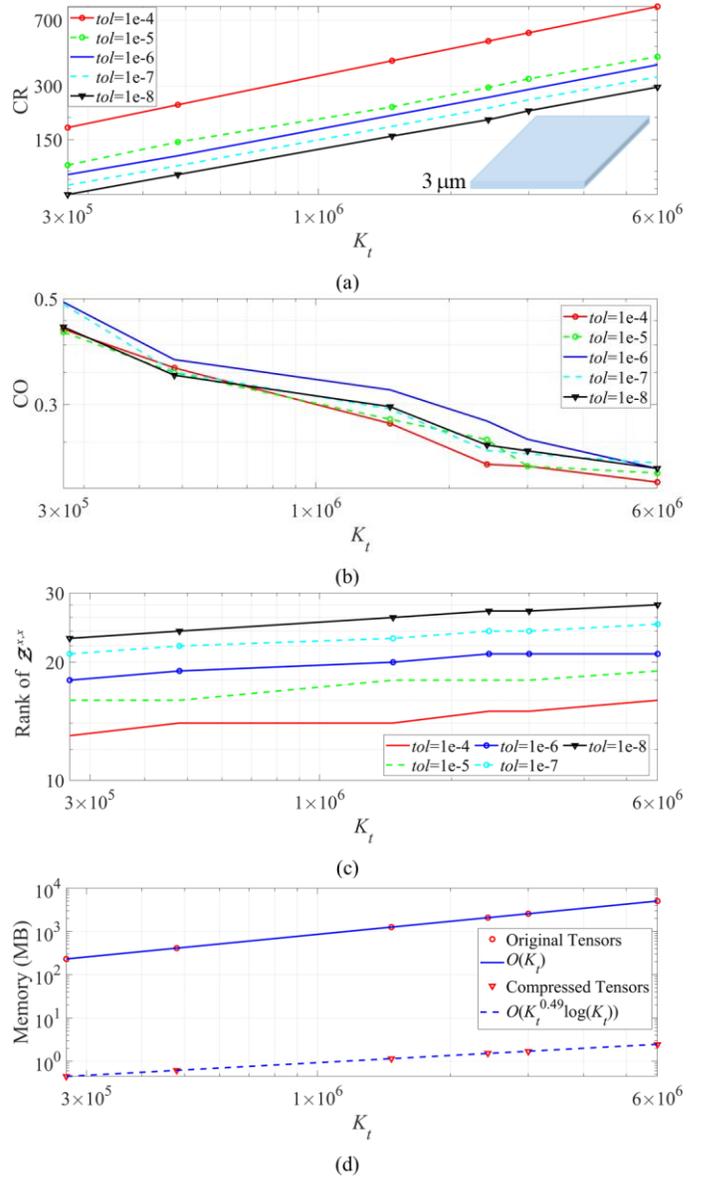

Fig. 2. The performance of the Tucker decomposition for compressing the circulant tensors generated for a 2D-like structure: (a) CR achieved by and (b) CO introduced by Tucker enhancement with increasing $K_t$. (c) The maximum multilinear rank of $\mathcal{Z}^{x,x}$ and (d) the memory scaling of original and compressed circulant tensors with respect to $K_t$.

Next, a 3D superconducting cube [Fig. 3(a)] is considered for analyzing the performance of Tucker decomposition on 3D structures. The edge length of the cube is set to $\{64, 100, 120, 450, 480\}\,\mu m$. For each different edge length, the structure is discretized by voxel of size $\Delta x = 1\,\mu m$. Fig. 3(a) shows that CR increases with increasing $tol$ and $K_t$; it reaches over 20,000 (for $tol = 10^{-4}$) for the largest structure. The CR achieved for the 3D cube is much larger than that achieved for the 2D-like structure due to larger circulant tensors generated for 3D cube. Fig. 3(b) shows that CO decreases with increasing $K_t$ and goes below 0.2 for the largest structure. Again, the maximum multilinear ranks of block $\tilde{\mathcal{Z}}^{x,x}$ for different $tol$ values are nearly constant with respect to $K_t$ [Fig. 3(c)]. Furthermore, the compressed circulant tensors' memory requirement scales as $O(K_t^{0.25} \log K_t)$, while the scaling of original tensors' memory requirement is $O(K_t)$ [Fig. 3(d)].

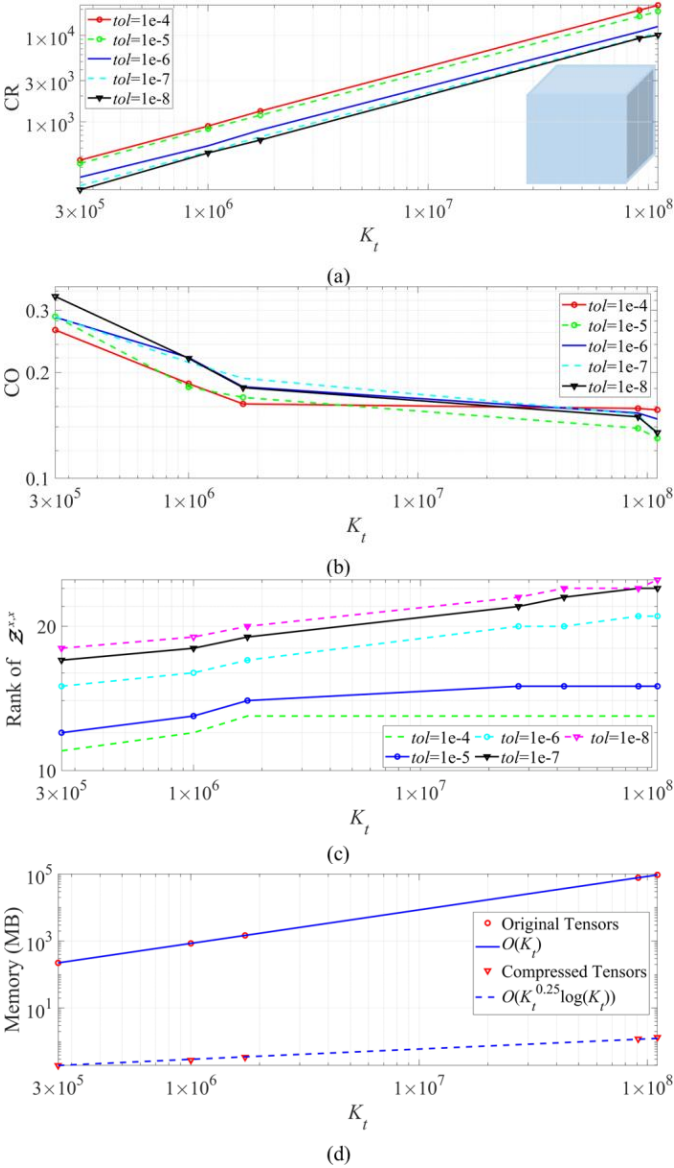

Fig. 3. The performance of the Tucker decomposition for compressing the circulant tensors generated for a 3D cube: (a) CR achieved by and (b) CO introduced by Tucker enhancement with increasing $K_t$. (c) The maximum multilinear rank of $\mathcal{Z}^{x,x}$ and (d) the memory scaling of original and compressed circulant tensors with respect to $K_t$.

*2) Compression of Toeplitz tensors:* The performance of Tucker enhancement for compressing the Toeplitz tensors and accelerating setup stage is presented here. To this end, edge length of a superconducting cube is increased from $100\,\mu m$ to $500\,\mu m$ with a step of $100\,\mu m$. For each different edge length, the cube is discretized with voxels of size $\Delta x = 1\,\mu m$ and the Toeplitz tensors are obtained directly. Furthermore, the Toeplitz tensors are obtained by the scheme outlined in Section II.B.1 (i.e., by reading Tucker-compressed tensors from hard disk, decompressing them, resizing them, and multiplying them with scaling factor). Table II shows that the memory of Tucker-compressed Toeplitz tensors is around MBs, negligible compared to the GBs of memory required by original Toeplitz tensors. It is apparent from the 3rd and 4th columns of the table that the proposed scheme for obtaining Toeplitz tensors requires negligible time compared to the time required for directly obtaining these tensors via computation. For the largest structure considered, the proposed scheme achieves 43.6x speed-up compared to the direct computation scheme for the Toeplitz tensors.

TABLE II
MEMORY AND CPU TIME REQUIREMENT FOR OBTAINING TOEPLITZ TENSORS BY DIRECT COMPUTATION AND FROM TUCKER-COMPRESSED TENSORS

| Edge length of cube ($\mu m$) | Memory of original Toeplitz tensors (MB) | Memory of Tucker-compressed tensors (MB) | CPU time to generate original Toeplitz tensors (s) | CPU time to obtain Toeplitz tensors from compressed tensors (s) |
|---|---|---|---|---|
| 100 | 106.81 | 1.95 | 31.78 | 0.527 |
| 200 | 854.49 | 2.96 | 86.20 | 1.14 |
| 300 | 2883.90 | 3.95 | 231.11 | 3.04 |
| 400 | 6835.94 | 5.37 | 535.09 | 13.06 |
| 500 | 13351.44 | 5.38 | 1042.26 | 23.92 |

*C. A 90-degree bend*

In this example, a 90-degree sharp bend with a sharp inner edge and a sharp outer corner is considered [Fig. 4(a)]. The bend with square cross-section ($2 \times 2\,mm$) consists of arms with lengths of $10\,mm$. The inductance of the bend is computed via SuperVoxHenry and FastHenry by setting $\lambda = 4.5 \times 10^{-4}\,m$ and $\sigma_0 = 0$. Initially, the structure is discretized by the voxels with $\Delta x = 0.025\,mm$ to obtain a reference inductance value. The accuracy of this reference value is checked by computing inductance obtained by one-level coarse discretization with $\Delta x = 0.05\,mm$. The relative difference between inductances obtained by SuperVoxHenry for $\Delta x = 0.025\,mm$ and $\Delta x = 0.05\,mm$ is computed as $8.4 \times 10^{-4}$; this validates the accuracy of the reference value. Next, the SuperVoxHenry and FastHenry with 2nd-order expansion are executed to obtain the inductances of the bend discretized with $\Delta x = \{0.5, 0.4, 0.25, 0.2, 0.1\}\,mm$. The accuracy of the inductance values is determined by computing their relative differences compared to the reference value. For the given relative difference, the memory requirement and CPU time of both simulators are plotted in Figs. 4(a) and (b). Clearly, SuperVoxHenry requires much less computational resources compared to FastHenry to achieve the same level of accuracy. For an accuracy level around $10^{-2}$, SuperVoxHenry requires 13x less memory resources and 876x less computational time compared to FastHenry. Furthermore, the current distribution on the bend discretized with $\Delta x = 0.1\,mm$ is provided in Fig. 4(b). For the same discretization, the memory and CPU time requirements of the FastHenry and SuperVoxHenry with and without Tucker enhancement are provided in Table III. Apparently, Tucker enhancement reduces the memory requirement 20% while introducing a small computational penalty of 9.41%. Table IV compares the performances of LDLT decomposition and AGMG method for inverting the Schur complement obtained while analyzing the structure with $\Delta x = 0.1\,mm$ and

$\Delta x = 0.05$ mm. For the bend discretized with $\Delta x = 0.1$ mm and $\Delta x = 0.05$ mm, AGMG method requires 7.69x and 12.09x less peak memory compared to the LDLT decomposition, respectively. At the same time, the AGMG method necessitates 2.79x and 11.62x less setup time to obtain the preconditioner compared to the LDLT setup time to obtain the decomposition. On the other hand, one inversion of the Schur complement via the AGMG method requires 6x and 2.45x more CPU time compared to the LDLT method. For the large-scale analysis, the AGMG is highly preferable compared to the LDLT decomposition technique since it yields significant memory reduction while imposing an acceptable computational penalty. For the analysis with $\Delta x = 0.05$ mm, the memory requirement of the LDLT decomposition of the Schur complement is 87.6% of the simulator's memory requirement. Reducing this requirement by a factor of 12.09x via the AGMG method boosts the proposed simulator's applicability to the analyses of large-scale structures.

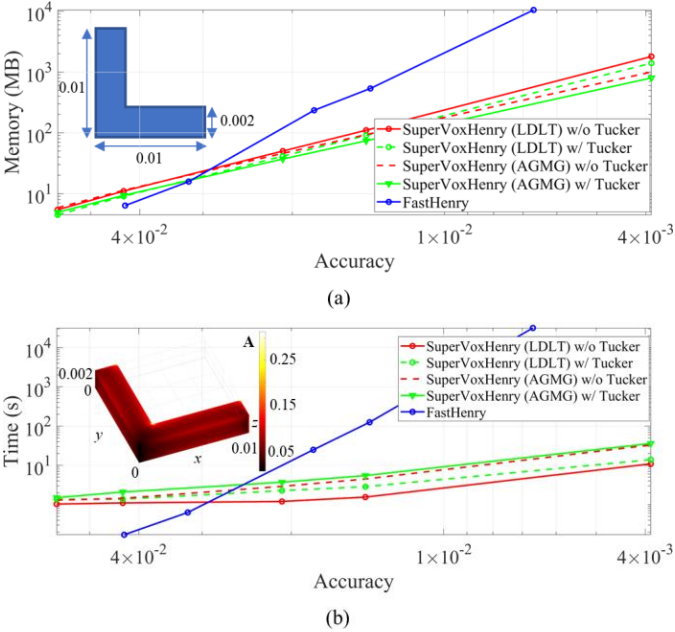

Fig. 4. (a) The memory and (b) CPU time requirements of the SuperVoxHenry and FastHenry simulators with respect to the accuracy of the inductance obtained for 90-degree bend with different discretizations.

TABLE III
MEMORY AND CPU TIME REQUIREMENTS OF SUPERVOXHENRY AND FASTHENRY FOR THE INDUCTANCE EXTRACTION OF 90-DEGREE BEND DISCRETIZED WITH $\Delta x = 0.1$ mm

|  | Memory (MB) | CPU Time (s) | Relative Difference |
|---|---|---|---|
| SuperVoxHenry w/o Tucker | 1,000 | 32.93 | 0.0039 |
| SuperVoxHenry w/ Tucker | 800 | 36.03 | 0.0039 |
| FastHenry w/ 2nd-order expansion | 10,500 | 31,569.41 | 0.0067 |

TABLE IV
MEMORY AND CPU TIME REQUIREMENTS OF AGMG AND LDLT DECOMPOSITION TECHNIQUES FOR INVERTING SCHUR COMPLEMENT

| $\Delta x$ (mm) | Technique | Peak Memory (MB) | Setup Time (s) | Inversion Time (s) |
|---|---|---|---|---|
| 0.1 | LDLT | 1,047 | 2.32 | 0.24 |
|  | AGMG | 136 | 0.83 | 1.44 |
| 0.05 | LDLT | 26,600 | 68.31 | 6.64 |
|  | AGMG | 2,200 | 5.88 | 16.29 |

### D. 10-bit eSFQ shift register

Finally, a 10-bit eSFQ shift register [1], realized by yttrium barium copper oxide (YBCO) with $\lambda = 4.5 \times 10^{-7}$ m and $\sigma_0 = 0$, is considered. The structure is enclosed by a computational domain with dimensions of $700.5 \times 142.5 \times 22.5$ μm and is discretized by voxels with $\Delta x = 0.5$ μm. The analysis at is performed by SuperVoxHenry with $tol = 10^{-4}$ and $RRE = 10^{-4}$. The normalized current distribution on the structure obtained by the proposed simulator is shown in Fig 5. The detailed breakdown of memory and CPU usage of the SuperVoxHenry for the analysis is presented in Table V. Needless to say, Tucker enhancement reduces the memory requirement of the circulant tensors from 16.47 GB to 4.91 MB with a CR of 3354. In parallel, it increases the iterative solution time by 20.35%. Furthermore, the Tucker enhancement accelerates the generation of the Toeplitz matrices by a factor of 26.3. The peak memory requirement of the AGMG method is less than one-fifth of that of the LDLT method. On the other hand, AGMG requires 4.38x more computational time compared to the LDLT for one Schur complement inversion. It is worthwhile to state here that the FastHenry can not be applied to this analysis since its memory requirement exceeds the computational resources used during this study. Furthermore, the performance of the proposed simulator on YBCO with isotropic superconductivity is the same as that on YBCO with anisotropic superconductivity.

TABLE V
DETAILED BREAKDOWN OF THE MEMORY AND CPU TIME USAGE FOR THE 10-BIT eSFQ SHIFT REGISTER EXAMPLE. UNITS FOR MEMORY AND CPU TIME ARE MB AND S, RESPECTIVELY.

| $N$ (# unknowns) | 47,161,696 |
|---|---|
| $K_t$ (# of voxels in comp. domain) | 5,816,192 |
| Memory for the incidence matrix | 1,641.84 |
| Memory for $\mathcal{Z}^{\beta,\alpha}$ (w/o Tucker) | 16,469.58 |
| Memory for $\mathcal{Z}^{\beta,\alpha}$ (w/ Tucker) | 4.91 |
| Memory for Schur complement | 3,075.10 |
| Peak memory for LDLT decomposition of Schur complement | 114,300 |
| Peak memory for AGMG inversion of Schur complement | 22,500 |
| Time for obtaining $\mathcal{T}^{\beta,\alpha}$ (via direct comp.) | 3,074.38 |
| Time for obtaining $\mathcal{T}^{\beta,\alpha}$ (via Tucker) | 116.92 |
| Time for compression of $\mathcal{Z}^{\beta,\alpha}$ | 137.73 |
| Time for the iterative solver (w/o Tucker) | 16,712.82 |
| Time for the iterative solver (w/ Tucker) | 20,113.87 |
| Number of GMRES iterations | 96 |
| Time for one inversion with LDLT | 44.03 |
| Time for one inversion with AGMG | 192.76 |

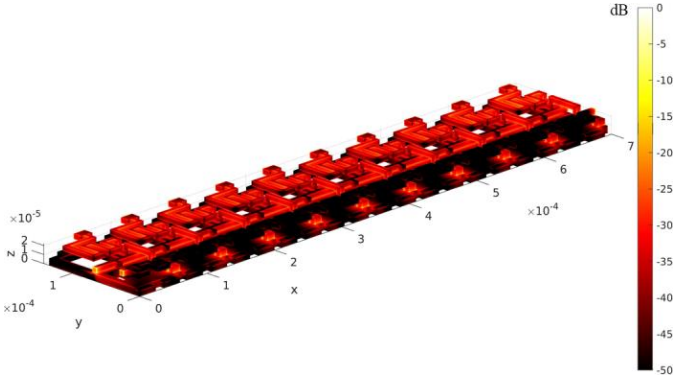

Fig. 5. The normalized current distribution on the 10-bit eSFQ shift register in the log scale.

## IV. CONCLUSION

In this paper, SuperVoxHenry, a Tucker-enhanced and FFT-accelerated inductance extraction simulator for voxelized superconducting structures, was proposed. The proposed simulator solves the VIE and the current conservation law after incorporating the two-fluid model. During the iterative solution of the VIE and current conservation law, the SuperVoxHenry expedites the matrix-vector multiplications via FFTs. To ensure the iterative solution's fast convergence, it uses a sparse preconditioner which is efficiently obtained by the AGMG method. Furthermore, it employs the Tucker decompositions to drastically reduce the memory requirement and setup time of the simulator. The numerical results showed that the proposed SuperVoxHenry simulator is far more memory and CPU efficient than the FastHenry for the inductance extraction of the voxelized superconducting structures. Needless to say, the proposed simulator is the memory and CPU efficient extension of VoxHenry and thereby useful for inductance extraction of the conducting structures as well.

## APPENDIX A
### THE ENTRIES OF DIAGONAL BLOCKS IN $\bar{\mathbf{Z}}$

With the incorporation of two-fluid model, the diagonal entries of the diagonal blocks (i.e., $l = k$) are computed via

$$\bar{\mathbf{Z}}_{lk}^{x,x} = \frac{a(\mathbf{r})}{\Delta x} + j\omega\mu\left(\frac{b(\mathbf{r})}{\Delta x} + \frac{1}{(\Delta x)^4}I_{lk}^1\right), \quad (16)$$

$$\bar{\mathbf{Z}}_{lk}^{2D,2D} = \frac{a(\mathbf{r})}{6\Delta x} + j\omega\mu\left(\frac{b(\mathbf{r})}{6\Delta x} + \frac{1}{(\Delta x)^6}I_{lk}^2\right), \quad (17)$$

$$\bar{\mathbf{Z}}_{lk}^{3D,3D} = \frac{a(\mathbf{r})}{2\Delta x} + j\omega\mu\left(\frac{b(\mathbf{r})}{2\Delta x} + \frac{1}{(\Delta x)^6}I_{lk}^3\right), \quad (18)$$

where

$$I_{lk}^1 = -\int_{V_l}\int_{V_k'} G(\mathbf{r},\mathbf{r}')\, dV'dV, \quad (19)$$

$$I_{lk}^2 = -\left(\int_{V_l}(x-x_l)\int_{V_k'}(x'-x_k)\, G(\mathbf{r},\mathbf{r}')\, dV'dV \right. \\ \left. +\int_{V_l}(y-y_l)\int_{V_k'}(y'-y_k)\, G(\mathbf{r},\mathbf{r}')\, dV'dV\right), \quad (20)$$

$$I_{lk}^3 = -\left(\int_{V_l}(x-x_l)\int_{V_k'}(x'-x_k)\, G(\mathbf{r},\mathbf{r}')\, dV'dV \right. \\ +\int_{V_l}(y-y_l)\int_{V_k'}(y'-y_k)\, G(\mathbf{r},\mathbf{r}')\, dV'dV \\ \left. +4\int_{V_l}(z-z_l)\int_{V_k'}(z'-z_k)\, G(\mathbf{r},\mathbf{r}')\, dV'dV\right), \quad (21)$$

$$a(\mathbf{r}) = \frac{\sigma_0(\mathbf{r})\left(\omega\mu\lambda(\mathbf{r})^2\right)^2}{1+\left(\sigma_0(\mathbf{r})\omega\mu\lambda(\mathbf{r})^2\right)^2}, \quad (22)$$

$$b(\mathbf{r}) = \frac{\lambda(\mathbf{r})^2}{1+\left(\sigma_0(\mathbf{r})\omega\mu\lambda(\mathbf{r})^2\right)^2}. \quad (23)$$

The off-diagonal entries of the diagonal blocks are the same as those given in [18] and $\bar{\mathbf{Z}}_{lk}^{x,x} = \bar{\mathbf{Z}}_{lk}^{y,y} = \bar{\mathbf{Z}}_{lk}^{z,z}$.

## APPENDIX B
### SCALING FACTOR

To obtain the Toeplitz tensors $\mathcal{T}^{\beta,\alpha}$, $\beta,\alpha \in \{x,y,z,2D,3D\}$, in a fast manner, these tensors are computed by setting $\Delta x = 1$ for a large computational domain, compressed by Tucker decomposition, and stored in the hard disk during the installation stage of the simulator. When executing the simulator, the Tucker-compressed tensors are read from the hard disk, restored to their original format, and resized with respect to the given computational domain. Then, the resized Toeplitz tensors are multiplied by the scaling factor, obtained from the voxel size. The scaling factor is derived by the change of variables so that integrals evaluated to compute the tensor entries become independent from $\Delta x$. Consider the integrals in (20)-(26) of the Appendix of [15]. Here we just derive the scaling factor for the integral pertinent to piecewise constant basis and testing functions (i.e., (20) in [15]). That said, the scaling factors of all integrals are the same as the derived scaling factor. The integral pertinent to piecewise constant basis and testing functions (i.e., $\beta,\alpha = \{x\}$) is

$$I^{x,x} = \int_{z_{s1}}^{z_{e1}}\int_{y_{s1}}^{y_{e1}}\int_{x_{s1}}^{x_{e1}}\int_{z_{s2}}^{z_{e2}}\int_{y_{s2}}^{y_{e2}}\int_{x_{s2}}^{x_{e2}} G_1\, dx'dy'dz'dxdydz, \quad (24)$$

where

$$G_1 = \frac{1}{4\pi\sqrt{(x-x')^2+(y-y')^2+(z-z')^2}}, \quad (25)$$

The integration on the primed coordinates $(x',y',z')$ is performed on the source volume (top voxel in Fig. A) for the basis function. The integration on the unprimed coordinates $(x,y,z)$ is evaluated on the observer volume (bottom voxel in Fig. A) for the testing function. In the voxelized setting, the source and observer voxels locate on grid points with indices $(n_x',n_y',n_z')$ and $(n_x,n_y,n_z)$. To this end, first, the integration intervals are changed as $x_{s2} = n_x'\Delta x$, $x_{e2} = (n_x'+1)\Delta x$, $y_{s2} = n_y'\Delta x$, $y_{e2} = (n_y'+1)\Delta x$, $z_{s2} = n_z'\Delta x$, $z_{e2} = (n_z'+1)\Delta x$, $x_{s1} = n_x\Delta x$, $x_{e1} = (n_x+1)\Delta x$, $y_{s1} = n_y\Delta x$, $y_{e1} = (n_y+1)\Delta x$,

$z_{s1} = n_z \Delta x$, and $z_{e1} = (n_z+1)\Delta x$. Then the variables and constants are rewritten as $x' = a'\Delta x$, $y' = b'\Delta x$, $z' = c'\Delta x$, $x = a\Delta x$, $y = b\Delta x$, and $z = c\Delta v$ in (24), which yields

$$I^{x,x} = \Delta x^5 \int_{n_z}^{n_z+1}\int_{n_y}^{n_y+1}\int_{n_x}^{n_x+1}\int_{n'_z}^{n'_z+1}\int_{n'_y}^{n'_y+1}\int_{n'_x}^{n'_x+1} G_2 \, da' db' dc' da \, db \, dc, \quad (26)$$

where

$$G_2 = \frac{1}{4\pi\sqrt{(a-a')^2 + (b-b')^2 + (c-c')^2}}. \quad (27)$$

The scaling factor is found to be $\Delta x^5$.

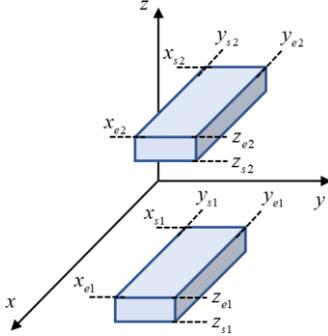

Fig. A. The sketch showing the geometrical quantities used for evaluating the integrals pertinent to piecewise constant basis functions.

electromagnetic design of complex systems in large fusion devices," *Plasma Phys. Control. Fusion,* vol. 63, no. 2, Dec. 2020.

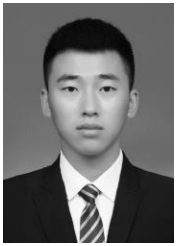

**Mingyu Wang** received the B.S. degree in electrical engineering and automation from Northeast Forestry University, China, in 2016 and M.S. in electronics from Nanyang Technological University, Singapore, in 2018. He is currently working toward the Ph.D. degree at the school of Electrical and Electronic Engineering, Nanyang Technological University, Singapore and works in Computational Electromagnetics Group.

His research is focused on computational electromagnetics and fast parameter extraction of voxelized interconnects & circuits.

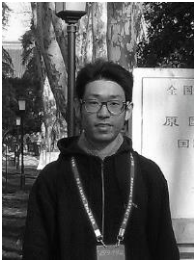

**Cheng Qian** received the B.S. and Ph.D. degree in electronics engineering from Nanjing University of Science and Technology, Jiangsu, China, in 2009 and 2015. From 2016 to 2018, he was a Research Associate with the Department of Applied Physics, The Hong Kong Polytechnic University, Hong Kong. Since 2018, he has been a Post-Doctoral Researcher with the School of Electrical and Electronic Engineering, Nanyang Technological University, Singapore. His current research interests include computational electromagnetics and nonlinear plasmonics.

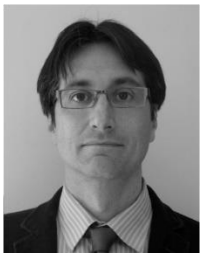

**Enrico Di Lorenzo** received the M.S. degree in electronics engineering (Summa Cum Laude) from Polytechnic University of Milan, Italy, in 1998, with a thesis on the electromagnetic characterization of ultra-miniaturized packages for Flash memories, developed under a stage by STMicroelectronics. From 1998 to 2000 he worked on robotics in the aerospace private sector, then joined Alcatel where he worked in the design of very high-speed digital signal systems. In 2006 he co-founded FastFieldSolvers S.R.L., maintaining and distributing open-source free solvers to the electromagnetic community, providing support, customization and simulation services.

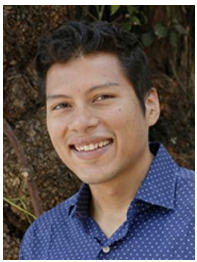

**Luis J. Gomez** (M'08) received the B.S. degree in electrical engineering and mathematics from the University of Florida, Gainesville, FL, USA, in 2008, and the M.S. degree in electrical engineering and applied mathematics and the Ph.D. degree in electrical engineering from the University of Michigan, Ann Arbor, MI, USA, in 2014 and 2015, respectively.

He is currently an Assistant Professor at Purdue University in the Elmore Family School of Electrical and Computer Engineering where he heads the SUBLIME lab. Previously, he was a Post-Doctoral Associate with the Department of Psychiatry and Behavioral Sciences, Duke University School of Medicine, Durham, NC, USA, where he developed optimization and computational techniques for use in improving noninvasive brain stimulation procedures. He was also a Post-Doctoral Fellow with the Radiation Laboratory, University of Michigan at Ann Arbor from 2015 to 2016, where he developed fast-integral equation methods for analyzing scattering by highly heterogeneous media and inverse scattering methods. His main research interests include computational electromagnetism, with a focus on wave propagation in highly heterogeneous materials, volume integral equations, electromagnetic imaging and optimization, and uncertainty quantification of biomedical applications.

Dr. Gomez was a recipient of the National Science Foundation Graduate Fellowship in 2008 and the National Institutes of Health BRAIN Initiative K99/R00 Post-Doctoral Training Award in 2019.

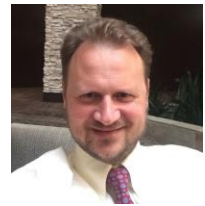

**Vladimir I. Okhmatovski** (M'99–SM'09) (M'99-SM'09) was born in Moscow, Russia, in 1974. He received the M.S. degree (with distinction) in radiophysics and Ph.D. degree in antennas and microwave circuits from the Moscow Power Engineering Institute, Moscow, Russia, in 1996 and 1997, respectively.

In 1997, he joined the Radio Engineering Department, Moscow Power Engineering Institute, as an Assistant Professor. He was a Post-Doctoral Research Associate with the National Technical University of Athens from 1998 to 1999 and with the University of Illinois at Urbana-Champaign from 1999 to 2003. From 2003 to 2004, he was with the Department of Custom Integrated Circuits, Cadence Design Systems, as a Senior Member of Technical Staff, and from 2004 to 2008 as an independent consultant. In 2004, he joined the Department of Electrical and Computer Engineering, University of Manitoba, Winnipeg, MB, Canada, where is currently a Full Professor. His research interests are the fast algorithms of electromagnetics, high-performance computing, modeling of interconnects, and inverse problems. He authored and co-authored over 150 technical papers, book chapters, and patents in the areas of computational and applied electromagnetics.

Since 2017, Prof. Okhmatovski has served on the Technical Program Review Committee (TPRC) of the IEEE Microwave Theory and Techniques Society (IEEE MTT-S) International Microwave Symposium (IMS), and since 2020 on IEEE MTT-S Technical Committee on Field Theory and Computational Electromagnetics. He was a TPC Co-Chair of 2021 Applied and Computational Electromagnetics Society Symposium.

Prof. Okhmatovski has been an active volunteer for the IEEE Antennas and Propagation Society (IEEE AP-S) as well as IEEE MTT-S serving as a Chapter Chair of the IEEE Winnipeg Waves Chapter from 2006 to 2011 and as Chair and Co-Chair of the Membership and Benefits Committee of the IEEE AP-S since 2018.

He was a recipient of the 2017 Intel Corporate Research Council Outstanding Researcher Award, 1995 scholarship of the Government of Russian Federation and the 1996 scholarship of the President of the Russian Federation. He was the recipient of the 1996 Best Young Scientist Report of the VI International Conference on Mathematical Methods in Electromagnetic Theory. He was also a co-recipient of the Best Paper Award at the 3rd Electronic Packaging Technology Conference in 2001 and Outstanding ACES Journal Paper Award in 2007.

Prof. Okhmatovski is a Registered Professional Engineer in the Province of Manitoba, Canada.

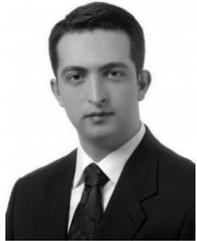

**Abdulkadir C. Yucel** received the B.S. degree in electronics engineering (Summa Cum Laude) from Gebze Institute of Technology, Kocaeli, Turkey, in 2005, and the M.S. and Ph.D. degrees in electrical engineering from the University of Michigan, Ann Arbor, MI, USA, in 2008 and 2013, respectively.

From September 2005 to August 2006, he worked as a Research and Teaching Assistant at Gebze Institute of Technology. From August 2006 to April 2013, he was a Graduate Student Research Assistant at the University of Michigan. Between May 2013 and December 2017, he worked as a Postdoctoral Research Fellow at various institutes, including the Massachusetts Institute of Technology. Since 2018, he has been working as an Assistant Professor at the School of Electrical and Electronic Engineering, Nanyang Technological University, Singapore.

Dr. Yucel received the Fulbright Fellowship in 2006, Electrical Engineering and Computer Science Departmental Fellowship of the University of Michigan in 2007, and Student Paper Competition Honorable Mention Award at IEEE International Symposium on Antennas and Propagation Symposium in 2009. He has been serving as an Associate Editor for the International Journal of Numerical Modelling: Electronic Networks, Devices and Fields and as a reviewer for various technical journals.